# Predicting Human Performance in Vertical Menu Selection Using Deep Learning

**Yang Li[1]**  **Samy Bengio[1,2]**  **Gilles Bailly[3]**

[1]Google Research & Machine Intelligence, Mountain View, CA, USA
[2]Google Brain, Mountain View, CA, USA
[3]Sorbonne Universités, CNRS, ISIR, Paris, France
{liyang, bengio}@google.com    yangli@acm.org    gilles.bailly@upmc.fr

## ABSTRACT
Predicting human performance in interaction tasks allows designers or developers to understand the expected performance of a target interface without actually testing it with real users. In this work, we present a deep neural net to model and predict human performance in performing a sequence of UI tasks. In particular, we focus on a dominant class of tasks, i.e., target selection from a vertical list or menu. We experimented with our deep neural net using a public dataset collected from a desktop laboratory environment and a dataset collected from hundreds of touchscreen smartphone users via crowdsourcing. Our model significantly outperformed previous methods on these datasets. Importantly, our method, as a deep model, can easily incorporate additional UI attributes such as visual appearance and content semantics without changing model architectures. By understanding about how a deep learning model learns from human behaviors, our approach can be seen as a vehicle to discover new patterns about human behaviors to advance analytical modeling.

## ACM Classification Keywords
H.5. HCI; I.2. Artificial Intelligence.

## INTRODUCTION
Seeking models for predicting human performance in performing an interaction task has long been pursued in the field of human computer interaction [2-8, 11, 16]. In addition to the scientific value of understanding human behaviors, creating these models has practical values in user interface design and development. A predictive model allows a developer or designer to understand the expected performance of an interface without having to test it with real users, which can be expensive and effort consuming.

Several predictive models of human performance have been devised, including Fitts' law [8] and Hick's law [11], which are rooted in information theory and experimental psychology. However, these models capture a certain aspect of human performance in isolation, e.g., motor control or decision making. They are limited in modeling human performance in realistic interaction tasks where multiple factors interplay. Recent work (e.g., [2]) has attempted to develop compound models that combine models such as Fitts' law. While these methods have made great progress in predicting time performance in more realistic tasks, these analytical models are not easily extensible to accommodate new factors that might come into play.

In this work, we take a departure from existing analytical approaches for performance modeling by using a data-driven approach based on the recent advance in deep learning [15]. Deep learning has proven successful in many domains, such as computer vision [15] and natural language processing [1]. It relieves the need of careful feature engineering and of a great amount of domain knowledge in creating a predictive model. It can also capture patterns that only manifest in the data but are difficult to articulate in an analytical form.

In particular, we devise a predictive model (see Figure 1) for interaction performance based on a deep recurrent neural net architecture using Long-Short Term Memory (LSTM) [12]. The unique architecture of our LSTM-based model allows us to naturally capture a variety of factors that come into play in UI tasks, including not only what human users are perceiving and performing at the moment but also what they have learned from the past regarding an interaction task. To scope our work, we focus on a common task on desktop and smartphones, where users select a target item from a vertical menu or list[1], e.g., choosing a song to play, a person to contact, selecting an application in the start menu or simply activating a command in a drop-down menu. Because users often need to perform these selection tasks repeatedly over the time, we investigate our approach in the context of a sequence of selection tasks.

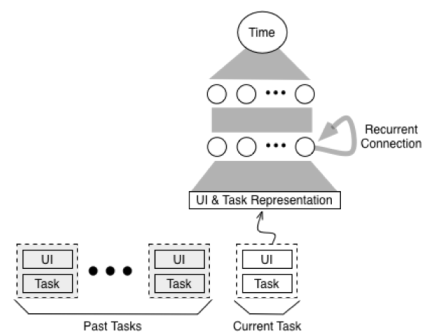

**Figure 1. A high-level illustration of our approach for predicting human time performance. The model, mimicking a human user, not only sees the UI and task at the moment but also remembers what it has experienced previously, through its recurrent layer. The representation of the UI and task at each step is achieved via another recurrent net.**

We design a novel hierarchical deep architecture for menu performance modeling. In our architecture, a recurrent neural net is used to encode UI attributes and tasks at each target item selection. This allows us to represent a menu with a varied length and to easily incorporate any additional UI attributes such as visual appearance and semantics. We then use another recurrent net to capture learning effects, a major component in human performance. The entire model is learned end-to-end using stochastic gradient descent. The model outperforms existing analytical methods in various settings for predicting selection time. Importantly, it is easily extensible for accommodating new UI features and human factors involved in an interaction task.

As a machine learning model, especially a deep architecture, the general challenge is that it is difficult to gain insights into what the model actually learns. We analyze how our model learns to mimic human behaviors. We show that our model "remembers" and "forgets" like a human—the model gains expertise on a visited item from past trials and that expertise fades away over time if the user does not access the item for a while. Prior work models expertise as frequency counts [2, 7], which does not take into account the "forgetting" effect in human behavior. We also discuss how this "memory effect" is affected by different menu organizations. We believe these analyses improve our understanding about how a deep learning model learns from human behaviors, which in turn can inspire analytical modeling.

**RELATED WORK**

Extensive work has been conducted in modeling human behaviors and predicting human performance for performing interaction tasks. For example, Fitts' Law [8] predicts the time needed for an expert human to acquire a visual target. Similarly, Hick's Law [11] is also a well-known model that describes the time required for an expert human to make a decision of choosing among a given number of options.

While each of these previous methods is amazingly robust for modeling the specific aspect of human behaviors it focuses on, they are limited in modeling realistic interaction tasks. For example, Fitts' Law was originally proposed for a limited setting of one-dimensional target with no distractors. Although prior work has extended Fitts' Law in several ways (e.g., [3, 16]), it is still constrained. Moreover, target acquisition is only one aspect of an interaction task. In a realistic task, there are many factors convoluted such as visual search, motor control and learning effects on spatial memory, as well as other factors that may or may not have been discovered in the literature.

There has been a considerable amount of efforts in combining these building block models such as Fitts' Law and Hick's Law so that their combination can be more applicable to modeling complex interaction tasks [2, 4, 6, 7, 10]. Particularly, GOMS/KLM [5] predicts the time taken for an expert user to perform a routine task. ACT-R [4] or EPIC [14] implement a set of production rules to decide, for instance, visual search strategies in linear menus. Prior work (e.g., [13, 21]) has also revealed the important role of learning effect played in menu selection, such as the forming of spatial memory reduces visual search.

More recently, Cockburn et al. [7] combined pointing time (Fitts' Law), decision time (Hick's Law), visual search time, and expertise in a single model. The model is compact and robust in modeling a range of menu selection tasks. In the same vein, Bailly et al. [2] proposed a more complex model that is formulated based on gaze distribution for menu selection tasks.

While previous methods, which are mostly empirically tuned analytical models, have gained substantial progress in modeling human behaviors, they are not easily extensible for accommodating various aspects of user interfaces and human factors. For example, the saliency of items [19] on an interface can significantly affect the time needed for visual search. Learning effect is a profound factor that affects every aspect of human performance. In addition, new generations of computing devices such as touchscreen smartphones have introduced many factors that are not covered by traditional models. While it is possible to further expand existing models, there is a tremendous amount of challenges and effort to do so—new factors are not always obvious or easily analytically articulated.

In contrast to previous approaches, we propose a data-driven approach based on the recent advance of deep learning. Deep learning [15] employs multiple processing layers to automatically learn representation of raw data, which reduces the need of intensive feature engineering and thus the need of domain expertise. In particular, we designed our model based on Long-Short Term Memory (LSTM) [12], a recurrent neural net, which has been successfully applied in many sequential problems such as natural language processing [1]. Our work is the first in applying deep recurrent neural net for modeling human performance in interaction tasks. While it is challenging to analyze the behavior of a deep model in general, we offer several insights into how such a model learns to mimic human behaviors from data.

---

[1] In this paper, we use "list" and "menu" interchangeably to refer to a list of items or options for the user to select.

**THE MODEL DESIGN & LEARNING**

We present a model to predict human performance in performing a sequence of UI tasks. It builds on two important capabilities of recurrent neural net (RNN) [9]. First, it is capable of "reading" in a varied-length sequence of information and encoding it as a fixed-length representation. It is important as an interaction task often involves varied-length information. For example, the number of items in a menu can vary from one application to another. Second, the model is capable of mimicking users' behavior by learning to both acquire and "remember" new experience, and discard (or "forget") what it learns if the experience is too dated. While learning effects are a major component in human performance, prior works primarily use a frequency count as the measure of the user's expertise. In contrast, our model relies on LSTM [12], which offers a mechanism that is more natural in mimicking human behaviors.

**Encoding A Single-Step Selection Task**

At each step, a user selects a target item in a vertical menu. From previous work, there are multiple factors in the task affecting human performance, including the number of items in the menu, the location of the target item in the menu, the visual salience of each item and whether there are semantic cues in assisting visual search.

For each element in the UI—an item in the menu in our context, we represent it as a concatenation of a list of attributes (see Equation 1). We use 1 or 0 to represent whether it is the target item for the current step. To capture the visual salience of an item, we use the length of the item name. An item that is especially short or long in comparison to the rest items on the menu tends to be easier to spot. To capture the semantics of an item, we represent the meaning of the item name with a continuous vector that was acquired from word2vec [20], which project a word onto a continuous vector space where similar words are close in this vector space. $m_s^j$ denotes the vector representation of the $j$th item in the menu at step $s$ in the interaction sequence.

$$m_s^j = [target, \text{len}(name), \text{word2vec}(name)] \quad (1)$$

To encode the selection task that involves a list of items in the menu, we feed the vector representation of each item in sequel to a recurrent neural net [9] (see Figure 2). $e_s^j$ represents the hidden state of the recurrent net after reading the $j$th item and seeing the previous items through $e_s^{j-1}$. $n$ denotes the number of items in the menu. This recurrent net performs as a task encoder (thereafter referred as the *encoder net*) and it does not have an output layer. The final hidden state of the recurrent net, $e_s^n$, a fixed-length vector, represents the selection task at step $s$. We then concatenate a one-hot vector to indicate whether the menu items are semantically grouped, alphabetically sorted or unsorted, resulting in $e_s$. The task encoder can accommodate a menu with any length, $n$, and UI attributes.

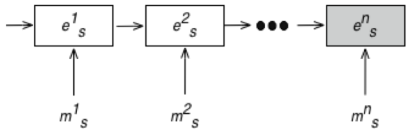

**Figure 2. We use a recurrent neural net to encode the selection task at a step, $s$. $m_s^j$ is the vector representation of the $j$th item in the menu and $e_s^j$ is the hidden state of the net after reading in the $j$th item.**

**Modeling A Sequence of Selection Tasks**

With the interaction task at each step of a sequence represented as $e_s$, we can now feed the sequence into another recurrent neural net (see Figure 3), which we refer to as the *prediction net*. Note that $e_s$ in Figure 3 represents the encoder net, which is a recurrent neural net itself whose outcome is fed to the prediction net. The task at each step can vary simply because the user might need to select a different target item. The UI at each step can also be different, e.g., an adaptive interface might decide to change the appearance of an item such as its size [7] to make it easier to acquire.

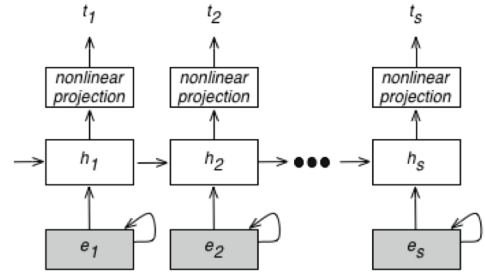

**Figure 3. A recurrent neural net takes in the selection task at each step, $i$, and the acquired expertise (represented as the hidden state $h_i$), and predicts the time needed, $t_i$, for completing the selection task at this step.**

The recurrent neural net predicts human performance time at each step, $t_i$. The predictions are based on not only the task at the current step but also the hidden state of the previous step that captures the human experience while performing previous tasks. Previous work in deep learning has shown that adding more layers in a deep net can improve the capacity for modeling complex behaviors [15]. To give the model more capacity, we add a hidden layer, with ReLU [18] as the activation function, after the recurrent layer, denoted as nonlinear projection in Figure 3. Finally, the time prediction $t_i$ is computed as a linear combination of the outcome of the nonlinear transformation layer.

**Model Learning & Loss Function**

It is straightforward to compute the time prediction with the feedforward process of a neural net. The two recurrent neural nets involved in our model are trained together, end to end from the data by feeding in sequences of selection tasks as input and observed performance times as the target output (the ground truth), using stochastic gradient descent.

For time performance modeling, one common measure of prediction quality in the literature has been $R^2$ (e.g., [2, 3, 7, 8]). It measures how well predicted times match observed ones in capturing relative task difficulty or human performance across task conditions and progression. For general time series modeling regarding continuous values, there are other metrics often used, such as Root Mean Square Error (RMSE) or Mean Absolute Error (MAE).

Mathematically, $R^2$ is the correlation between the sequence of observed times, $y_i$, and the sequence of predicted times, $t_i$, (see Equation 2). $|S|$ represents the length of the sequence, and $\bar{y}$ is the mean of $y_i$.

$$R^2 = 1 - \frac{\sum_{i=1}^{|S|}(y_i - t_i)^2}{\sum_{i=1}^{|S|}(y_i - \bar{y})^2} \quad (2)$$

$\sum_{i=1}^{|S|}(y_i - \bar{y})^2$ reflects the variance of the observations in each sequence, which is independent of models. Thus, it is a known constant for each sequence in the training dataset, which we refer to as $c_s$. To maximize $R^2$, we want to minimize the squared error term $\sum_{i=1}^{|S|}(y_i - t_i)^2$, scaled by a sequence-specific constant $c_s$, which defines the loss function (see Equation 3). The scaling acts effectively as adapting the learning rate based on the variance of each sequence for training the deep neural net. Intuitively, for each training sequence, the more variance the sequence has, the smaller learning rate we should apply for updating the model parameters, and vice versa.

$$L_t = \frac{\sum_{i=1}^{|S|}(y_i - t_i)^2}{c_s} \quad (3)$$

With the loss function defined, our model can be trained using Back Propagation Through Time (BPTT) [9], a typical method for training recurrent neural nets (see more details in the following sections).

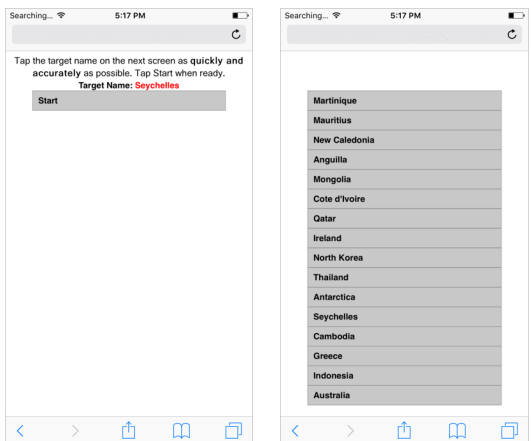

**Figure 4. The Data Collection web application on iPhone. For each trial, with informed the target name, the worker taps on the Start button to reveal the menu and then selects the target.**

## EXPERIMENTS

We experimented with our model based on two datasets from both a controlled and a crowd-sourced environment. We first discuss these datasets, and the model and training configuration. We then discuss the performance of our model in comparison with a state of art analytical method proposed previously [2].

### Datasets

We experimented with the deep net on modeling menu selection tasks with a public dataset [2] collected from a desktop computer in a controlled laboratory environment, and a dataset we collected from smartphone users using crowdsourcing. For testing different task difficulties and learning effects, these datasets involve menus with different lengths and long sequences of trials.

*Public Dataset from a Controlled Laboratory*

Bailly et al. conducted a within-subjects study to collect data from 21 participants [2]. The study involved three menu organizations: Unordered, Alphabetical and Semantic, and three menu lengths: 8, 12 or 16 items. Participants were asked to complete 12 blocks of trials per menu. For each menu, the order of items in the menu is fixed so that participants can learn item positions over time. Within a block, each target item is selected once and the order of target items was randomized. The experiment was conducted on a windows PC with a 20 LCD display and a traditional optical mouse. In total, there are 189 sequences: 21 participants x 3 Menu configurations x 3 Menu Length. The length of each sequence may vary depending on the menu length: 12 blocks x (8-12-16 items) and there are 39,564 selections.

*Touchscreen Smartphone Dataset from Crowdsourcing*

To collect data of users interacting on a smartphone, we implemented a data collection tool as a web application (see Figure 4). We recruited smartphone users via Amazon Mechanical Turk and these users are redirected to our web application to complete a sequence of menu selection tasks. Because the data collection takes place in the wild, we intend to maximally assure the consistence of the setting by controlling a few setups. To make sure a list to appear with the same size on a phone, we decide to target on a few specific smartphone models: Apple iPhone 6 and Android Nexus 5. A user is also required to hold the phone in the portrait mode while performing the task, and our web app automatically prohibits the users to continue the task if it detects the landscape mode. Once a user completed a sequence of tasks, the data are automatically uploaded and stored on the server.

Our data collection was designed in a similar way to previous work. There are three menu lengths: 8, 12 and 16. Each worker is randomly assigned to a menu length with the item labels also randomly selected from a set of country names. The order of items in each menu is randomly determined for each worker and remains fixed throughout the trials. The location of the target item in the menu is randomly assigned for each trial. The number of trials completed by each worker ranged from 96 to 192 depending on menu lengths. In total, there were 863 sequences generated, each from a unique smartphone user (804 iPhone and 59 Android users). There were 384 males and 479 females. 50% users aged between 20-29 and 32% users between 30-39. 10% of these users were left-handed. The mean duration of a worker session spanned 7 minutes (STD=2.6 minutes). There are in total 159,072 selections in these sequences.

**Model Configuration & Hyper Parameters**
For the encoder net, the recurrent layer uses 16 LSTM cells [12]. The text name of each menu item is represented as a continuous embedding vector based on a 50-dimensional vector representation learned from the Wikipedia corpus [20]. Because the number of unique names in our datasets is relatively small compared to the entire English vocabulary on Wikipedia, we reduce the dimensionality of the name embedding to 4 using PCA, to speed up learning. As a result, each menu is represented as a continuous vector of size 6: 1 slot for indicating if the item is the target, 1 slot for the name length, and 4 slots for the name embedding. For the prediction net, we used 32 LSTM cells for its recurrent layer. The nonlinear transformation over the recurrent layer has a size of 16 that are used for computing the time prediction. We implemented our model in Python based on TensorFlow, an open source deep learning library [17].

We trained the model by minimizing the loss in Equation 3 using Ada adaptive gradient descent, with a learning rate of 0.01, a norm of 1.0 for clipping the gradients, a batch size 1 and the number of unroll of 40 for truncated back-propagation through time [9]. To regularize the model learning, we found applying a dropout ratio of 10% to the task encoding, $e_s$, can effectively avoid the model overfitting the training data prematurely.

**Performance Results**
For each dataset, we randomly split the data on users with half of the users for training and the other half of the users for testing the model. For the experiments with the public dataset, our model was trained for four million iterations. Because there are significantly more data in the smartphone crowd dataset, we trained our model for ten million iterations.

Previous modeling work often reports performance on how well a model can fit the data. As a machine learning model that uses a large number of parameters, it is meaningless to report fitting performance. In this paper, we only report the experiment results based on the test dataset for which the model was not trained on, which truly shows how well the learned model can predict on new data.

We report the accuracy of our model on both target-level and menu-level $R^2$ that were used in previous work [2]. Both measure the correlation between predicted and observed performance times. Target-level $R^2$ examines performance at each target position in a menu with a different amount of practice (blocks). Menu-level $R^2$ examines the average performance over all target positions in a menu with a varying amount of practice.

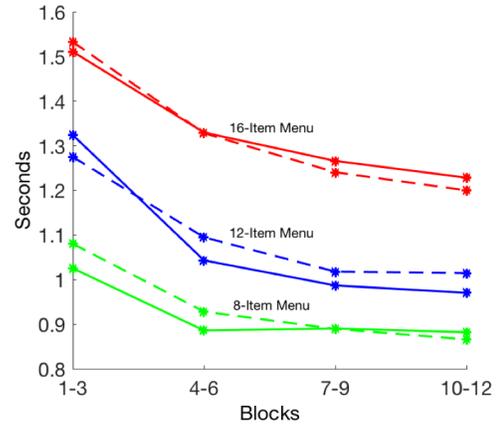

Figure 5. The model accuracy over blocks on the smartphone dataset for menus with different lengths, with predicted times in solid lines and observed times in dashed lines.

For target-level $R^2$, our model achieved 0.75 on the public dataset for the overall correlation across menu organizations. In particular, $R^2$ for alphabetically ordered (A), semantically grouped (S) and unsorted menus (U) are 0.78, 0.62 and 0.80 respectively. Note that we used a single model to predict for all menu organizations. Previously, Bailly et al. tuned and tested their model for each menu organization separately. Their $R^2$ results were reported as 0.64 (A), 0.52 (S), 0.56 (U) [2]. For menu-level $R^2$, our model achieved 0.87 for overall correlation: 0.85 (A), 0.88 (S) and 0.94 (U), which previous work reported as 0.91 (A), 0.86 (S) and 0.87 (U) [2]. Our model achieved a similar performance for the smartphone dataset that involves only unordered menus: target-level $R^2 = 0.76$ and menu-level $R^2 = 0.95$. It accurately predicts the time performance for each menu length (see Figure 5).

**ANALYZING MODEL BEHAVIORS**
While it is generally challenging to analyze what a deep model learns, we here offer several analyses of the model behaviors and discuss how they match our intuition about user behaviors. To understand the behavior of our deep net model, we compute its Jacobian that is the partial derivatives of the network output with respect to a specific set of inputs (see Figure 6)—it indicates how sensitive the time prediction is to the change in the input at each step. In particular, we want to find out how users' past experience with selecting a target affects their performance for selecting the target item again.

Figure 6 is generated by taking the Jacobian of the deep net output, i.e., the time performance, with regards to the *target* feature in Equation 1. We see that the more recent the experience is with selecting a target item, the more influence it has on the current trial for selecting the target item again. Intuitively, it might be because the user remembers where the item is on the list, as found in previous work [2, 7]. However, such an effect eventually wears off as the experience becomes dated, which is quite consistent with how human memory works. Previous work uses frequency count to represent the user expertise with an item and is insufficient to capture the profound aspects of human short-term memory such as the forgetting effect.

We also found the degree of how much the current performance relies on the past experience differs for the different menu organizations. As shown in Figure 6, such sequence dependency has the largest effect on unordered menus and less effect on semantically or alphabetically organized menus. We speculate that this is because semantic and alphabetic menus provide additional cues for users to locate an item, which result in less dependency on memory for completing the task.

## DISCUSSION & FUTURE WORK

LSTM has been extensively used in domains such as natural language processing (NLP). One major commonality between interaction tasks and NLP is that they both are concerned with sequential dependency, which is manifested as learning effect in interaction tasks. But unlike a standard NLP problem that deals with a sequence of words (tokens), each step in our selection task involves visual search and acquisition of the target in the percept of the menu UI with various attributes, which cannot be simply treated as a token. We designed the model architecture based on LSTM to capture this unique aspect of interaction tasks. The memory cells in LSTM is to simulate human short-term memory that is updated when the user performs each selection task. The execution of a task trial is simulated by another LSTM where visual search and motor control are captured. We found the memory capability is important for modeling human performance and our early exploration with a feedforward neural net did not yield good performance.

There are several advantages for using a deep learning approach for human performance modeling. Creating an effective analytical model requires a lot of domain expertise, including perception, motor control and decision making. Manually combining different performance components is hard to scale as more and more factors are involved in a realistic interface. In contrast, a deep learning approach is easily extensible for incorporating new factors as long as they are manifested in the data. Our model, which employs two hierarchically organized recurrent neural nets, is highly extensible to incorporate additional UI and task attributes.

That said, analytical models are easier to understand than a deep learning model. In fact, it has been a general challenge for the interpretability of deep learning models although they offer superior accuracy performance. In this work, we analyzed how human learning effects are captured and mimicked by a deep model (see Figure 6), and how the learning effect differs across different menu organizations. The Jacobian responds differently for different menu organizations, which reflects memory effect as manifested from the data. These findings can advance our understanding about how human behaves thus may inspire others to design new analytical models to capture these effects. Thus, rather than using deep models only for predictions, the deep learning approach can be seen as a vehicle to discover new patterns about human behaviors to advance analytical research.

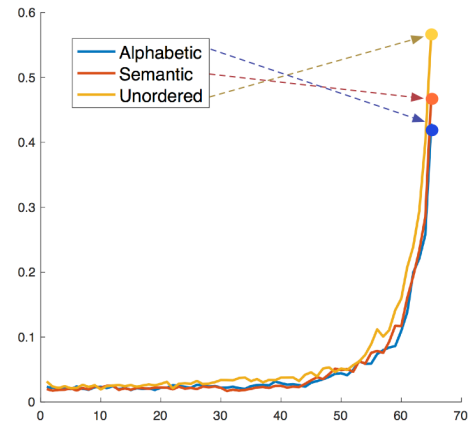

Figure 6. The Jacobian of our deep net indicates how the time performance for selecting a target item is affected by the past experience for selecting the item, in response to different menu organizations. The X axis is the trials in the sequence and the Y axis shows the magnitude of the derivative, i.e., the impact.

There are several directions for future work. It is worth examining the analytical findings about the memory effects on different menu organizations by conducting further user experiments. In addition, there are opportunities to extend our model for multilevel analysis of interaction by considering finer-granularity behaviors such as the mouse and gaze path of users in selection tasks. This has the potential to provide new insights on visual search in menus and the coordination between gaze and hand movements.

## CONCLUSION

We presented a deep learning approach for modeling user performance for menu selection, a dominant task in modern interfaces. Our model is highly extensible. It can accommodate various UI aspects without changing the model architecture or using extensive domain knowledge. It outperformed a previous method on predicting performance time based on a public dataset and a large-scale smartphone dataset. We discussed an analysis of the model behaviors, which revealed new findings about how past experience of the user has an influence on the user performance with regard to different menu organizations. We contributed a set of knowledge and technical details about how to design, train and analyze a deep model for performance modeling.